\documentclass{aa}
\usepackage{graphicx}
\begin{document}
\titlerunning{The 0.4-M$_{\sun}$ Eclipsing Binary CU Cnc}
   \title{The 0.4-M$_{\sun}$ Eclipsing Binary CU Cancri\thanks{Tables 1 and 2 
          are only available in electronic form at the CDS via anonymous 
          ftp to cdsarc.u-strasbg.fr (130.79.125.5) or via 
          {\tt http://cdsweb.u-strasbg.fr/Abstract.html}}}
   \subtitle{Absolute Dimensions, Comparison with Evolutionary Models and
             Possible Evidence for a Circumstellar Dust Disk}

   \author{I. Ribas}

   \offprints{I. Ribas}

   \institute{Departament d'Astronomia i Meteorologia, Av. Diagonal, 647,
              E-08028 Barcelona, Spain\\
              \email{iribas@am.ub.es}
             }

   \date{Received ; accepted}

   \abstract{Photometric observations in the $R$ and $I$ bands of the
detached M-type double-lined eclipsing binary \object{CU Cnc} have been
acquired and analysed.  The photometric elements obtained from the
analysis of the light curves have been combined with an existing
spectroscopic solution to yield high-precision (errors~$\la$2\%) absolute
dimensions: $M_{\rm A}=0.4333\pm0.0017$~M$_{\sun}$, $M_{\rm
B}=0.3980\pm0.0014$~M$_{\sun}$, $R_{\rm A}=0.4317\pm0.0052$~R$_{\sun}$,
and $R_{\rm B}=0.3908\pm0.0094$~R$_{\sun}$. The mean effective temperature
of the system has been estimated to be $T_{\rm eff}=3140\pm150$ K by
comparing multi-band photometry (optical and infrared) with synthetic
colors computed from state-of-the-art model atmospheres. Additionally, we
have been able to obtain an estimate for the age ($\sim$320 Myr) and
chemical composition ($[\mbox{Fe/H}]\approx 0.0$) of the binary system
through its membership of the Castor moving group. With all these
observational constraints, we have carried out a critical test of recent
stellar models for low-mass stars. The comparison reveals that most
evolutionary models underestimate the radius of the stars by as much as
10\%, thus confirming the trend observed by Torres \& Ribas (\cite{TR02})
for \object{YY Gem} and \object{V818 Tau}. In the mass--absolute magnitude
diagram, CU Cnc is observed to be dimmer than other stars of the same mass
and this makes the comparison with stellar models not so compelling. After
ruling out a number of different scenarios, the apparent faintness of CU
Cnc can be explained if its components are some 10\% cooler than
similar-mass stars or if there is some source of circumstellar dust
absorption. The latter could be a tantalizing indirect evidence for a
coplanar (Vega-like) dusty disk around this relatively young M-type
binary.

   \keywords{binaries: eclipsing --
             binaries: spectroscopic --
             Stars: low-mass, brown dwarfs --
             Stars: fundamental parameters --
             Stars: individual (CU Cnc) --
             planetary systems: protoplanetary disks
             }
            }

   \maketitle
%

\section{Introduction}

Low-mass stars constitute the most numerous stellar population of the
Galaxy. Yet, their physical properties are still poorly known. This is not
without the efforts of many investigators that have been working
intensively on the modelling of the atmosphere and interior structure of
sub-solar mass stars. The major problem that the field has to face is a
conspicuous lack of suitable calibrators. Unfortunately, very few
main-sequence M stars have empirically-determined masses, radii,
luminosities and temperatures.

Eclipsing binaries have often served as valuable benchmarks for the
validation of structure and evolution models (see, e.g., Andersen
\cite{A91}). Despite some unsolved problems at the high-mass end of the
main sequence, the behaviour of stars more massive than the Sun is well
understood, partly thanks to the large number of eclipsing binaries with
well-determined physical properties. The situation is radically different
at the low-mass end of the main sequence. Selection effects caused by the
faintness of the stars and the often strong intrinsic variations due to
magnetic activity pose a serious challenge for the discovery and analysis
of detached eclipsing binaries. Only three eclipsing systems with M-type
components have been identified to date. The member of the Castor multiple
system \object{YY Gem} has almost identical components of spectral type
M1~Ve and mass $\sim$0.6~M$_{\sun}$ (Torres \& Ribas \cite{TR02},
hereafter TR02). \object{CM Dra} holds the record for the lowest mass
eclipsing binary known and has components of $\sim$0.25~M$_{\sun}$ and
spectral type M4.5~Ve (Metcalfe et al. \cite{MML96}). These were the only
two M-type eclipsing binaries known for decades until Delfosse et al.
(\cite{Dea99a}, hereafter D99) recently reported the discovery of eclipses
in \object{CU Cnc}.

CU Cnc (GJ 2069A, HIP 41824, $\alpha=8^h 31^m 37\fs58$, $\delta=+19\degr
23\arcmin 39\farcs5$) is an 11.9-mag spectroscopic binary with M3.5~Ve
components. Also, it has a fainter visual companion, \object{CV Cnc} or GJ
2069B, at an angular distance of $\sim$12\arcsec. GJ 2069B is itself a
close binary star (Delfosse et al. \cite{Dea99b}) with a mean spectral
type of M4 and appears to form a physically bound quadruple system with CU
Cnc. Interestingly, the masses found by D99 place CU Cnc nicely in between
of YY Gem and CM Dra and make it a system of paramount importance to
improve upon our knowledge of the physics of low-mass stars.

Currently, the high-precision spectroscopic orbit of CU Cnc presented by
D99 (errors in minimum mass below 0.4\%) is not matched by the
photometric light curves, and the authors reported only sparse
photoelectric observations. The motivation of our study was to acquire and
analyse new high-quality photometric observations with the aim of
determining accurate masses and radii for the components of CU Cnc. With a
careful estimation of the temperature, our goal is to compare the physical
properties of CU Cnc's components with the predictions of the available
low-mass stellar evolution models.

\section{Light curve observations} \label{secLC}

The just recent discovery of the eclipsing nature of CU Cnc is likely to
be a consequence of its faintness but also, very importantly, of its
shallow and short eclipses. The photometric observations published by D99
indicate a system with an orbital period of 2.77 days, eclipses of only
$\sim$0.2 mag in depth and about 2 hours in duration. However, the high
scatter of these light curves and their poor phase coverage prevented an
accurate determination of the physical properties of CU Cnc.

We report here new photometry of CU Cnc in the Johnson $R$ and $I$ bands.
Both the photometric accuracy (a few millimagnitudes) and the phase
coverage (over 2100 observations) are sufficient to guarantee a reliable
determination of the light curve parameters thus permitting a critical
evaluation of stellar models. The observations were carried out with the
Four College 0.8-m APT, which is equipped with a refrigerated Hamamutsu
photoelectric detector and filters closely matching the standard Johnson
system. Differential photometry was obtained in which \object{HD~72093}
(F8, $V=7.80$) was employed as the comparison star, while
\object{HD~72358} (F5, $V=8.31$) served as the check star. Although it is
common practice to employ comparison stars that match the spectral type of
the variable star, we preferred to avoid G, K or M stars because of
concerns with magnetic-related intrinsic variability. In contrast, late F
stars are known to be photometrically very stable. Note that only precise
$R$ and $I$ photometry could be obtained with our instrumental setup
because of the faintness of CU Cnc at shorter wavelengths. The system is
so red that at the $I$ band, its brightness rises up to about $I\approx9$
mag. No evidence of significant light variations (down to the few
milimagnitude level) was found for the comparison--check star sets. The
observations were reduced using photometric reduction programs at
Villanova University (USA). Differential extinction corrections were
applied, although these corrections were typically very small.

\begin{figure*}
\centering
\includegraphics[width=\textwidth]{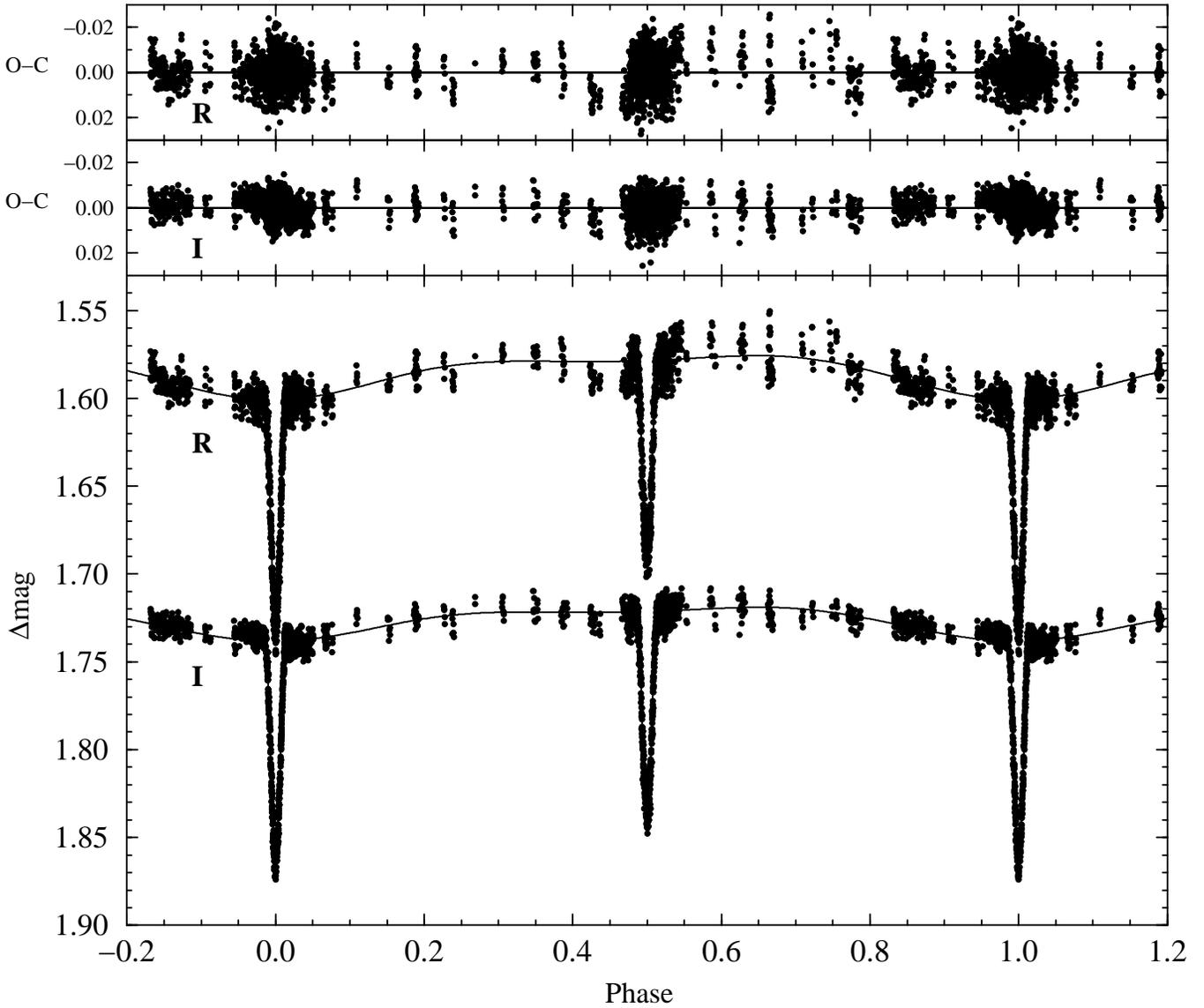}
\caption{$R$ and $I$ light curves of CU Cnc. A zero point shift of $-1.45$ has
been applied to the $R$ photometry for display purposes. The solid line in the
bottom panel is the best-fitting synthetic light curve generated by W-D. The
top panels show the residuals of the fits. Enlargements of the fits to the
eclipse phases are shown in Fig. \ref{figEcl}.}
\label{figLC}
\end{figure*}

There is one feature that makes our photometric dataset especially
noteworthy: the $\sim$2100 measurements were obtained over a period of
only 110 days (from December 5, 1998 through March 25, 1999). Classified
as a flare star, CU Cnc is expected to be magnetically active and display
some level of brightness variability due to surface inhomogeneities. Thus,
the photometric observations were acquired over a short time span to
minimise possible variations in the out-of-eclipse shape caused by
starspot migration. Also, special care was taken to schedule the
observations near the eclipses so that a very dense phase coverage could
be achieved. To compute orbital phases, we adopted the ephemeris of D99
but correcting the spectroscopic reference epoch to the primary eclipse:
\[\begin{array}{rcrcrc}
T(\mbox{Min I})&=&{\rm HJD}2450208.5068&+&2.771468 & E \\
\end{array}\]
The phased light curves are presented in Fig. \ref{figLC}. Note the dense
phase coverage, the relative shallowness of the eclipses and the
out-of-eclipse variations presumably due to starspots. The individual
photometric observations in the $R$ ($n=2157$) and $I$ ($n=2188$) bands
are provided in Tables 1 and 2 (HJD, phase and $\Delta$mag), respectively,
which are only available in electronic form. \addtocounter{table}{2}

As mentioned above, CU Cnc has an optical companion at $\sim$12\arcsec.
The diaphragm used for the photometric observations was 45\arcsec\ in
diameter, so the light of the companion was included in the measurements.
To estimate exactly the fraction of contributed light we made use of CCD
images taken with the ALFOSC instrument at the Nordic Optical Telescope
(La Palma). Several images in the $V$ and $I$ passbands were kindly
acquired by R. J. Irgens on May 10, 1999 (HJD 2451309.375) at an orbital
phase of 0.215. The reduction was carried out in the classical manner and
aperture photometry was performed to estimate the differential magnitude
between CU Cnc and its companion. We obtained values of $\Delta
V=1.45\pm0.01$ and $\Delta I=1.20\pm0.01$, which translates into the
following fractions of 3rd light, that can be directly incorporated into
the analysis of the light curves: \[F_3^{\rm V} = 0.21\pm0.01;
\hspace{3mm} F_3^{\rm I} = 0.25\pm0.01\] From these we estimate the
fraction of 3rd light in the $R$ passband to be $F_3^{\rm R} =
0.23\pm0.01$.

\section{Light curve analysis} \label{secSol}

The fits to the light curves were performed using an improved version of
the Wilson-Devinney program (Wilson \& Devinney \cite{WD71}; hereafter
W-D) that includes a model atmosphere routine developed by Milone et al.
(\cite{MSK92}) for the computation of the stellar radiative parameters. A
detached configuration was chosen for all solutions. Both reflection and
proximity effects were taken into account, even though the light curves do
not show strong evidence for these.  The bolometric albedo was set at the
canonical value of 0.5 for stars with convective envelopes. The
gravity-brightening coefficient was adopted as 0.2 following the
theoretical results of Claret (\cite{C00a}). A mass ratio of
$q=M_B/M_A=0.9184$ was adopted from D99 and the temperature of the primary
component (eclipsed at phase 0.0) was set to 3160~K (see Sect.
\ref{secAbs} for discussion). For the limb darkening we used a logarithmic
law as defined in Klinglesmith \& Sobieski (\cite{KS70}). In our W-D
implementation, first- and second-order limb darkening coefficients are
interpolated at each iteration from a set of tables computed in advance
using a grid of Kurucz model atmospheres. However, CU Cnc has a
temperature somewhat below the coolest model and a extrapolation had to be
made. We also ran some tests by fixing the coefficients at the values
computed by Claret (\cite{C00b}) for the appropriate effective
temperatures. The change had negligible effects but we finally adopted
this latter prescription to avoid extrapolations. Finally, third light
fractions were set to the values provided in Sect. \ref{secLC}.

The iterations with the W-D code were carried out automatically until
convergence, and a solution was defined as the set of parameters for which
the differential corrections suggested by the program were smaller than
the internal probable errors on three consecutive iterations. As a general
rule, several runs with different starting parameters are used to make
realistic estimates of the uncertainties and to test the uniqueness of the
solution.

Simultaneous W-D fits to the $RI$ light curves were carried out using the
mean error of a single measurement as the relative weight, $w_{\lambda}$,
of each passband (``curve-dependent'' weighting scheme). Note that the
number of photometric measurements we have is so large that 4-point
normals had to be computed at the densest phases in or around the eclipses
when running the W-D fits. We initially solved for the following light
curve parameters: the orbital inclination ($i$), the temperature of the
secondary (${T_{\rm eff}}_{\rm B}$), the gravitational potentials
($\Omega_{\rm A}$ and $\Omega_{\rm B}$), the luminosity of the primary at
each passband ($L_{\rm A}$), a phase offset ($\Delta\phi$), and the spot
parameters. The orbital eccentricity was set to zero as the light curves
do not show any evidence for eclipses of different width or a secondary
eclipse at an orbital phase other than 0.5. Some tests were run to check
for the possibility of a small non-zero orbital eccentricity. The W-D
solutions converged towards a value of $e=0.001\pm0.003$, thus consistent
with our adoption of a null eccentricity.

As expected for a system with partial eclipses and similar components,
numerous test solutions revealed that the ratio of the radii ($k\equiv
r_B/r_A$, where $r_A$ and $r_B$ are the fractional radii in units of the
separation) is poorly constrained. The problem is, in principle, severe
because it implies that the light curves are not sufficiently sensitive to
discriminate between the sizes of the components\footnote{Note, however,
that the sum of fractional radii is extremely well determined because it
depends strongly on the total duration of the eclipses.}. Thus, some
source of external information has to be used to infer a value for $k$.
With the ratio of temperatures well determined from the light curves, a
spectroscopic estimation of the luminosity ratio would provide the
necessary information.

Two high resolution and $S/N$ echelle spectra were obtained with the UES
instrument at the William Herschel Telescope (La Palma) as part of the
Service Programme. The observations were carried on March 8--9, 2001 (HJD
2451977.499) at orbital phase 0.287. The raw images were reduced using
standard NOAO/IRAF tasks (including bias subtraction, flat field
correction, sky-background subtraction, cosmic ray removal, extraction of
the orders, dispersion correction, merging, and continuum normalization).
Equivalent width ratios for the components were measured for 10 relatively
clean and isolated Ca~{\sc i} and Fe~{\sc i} absorption features. The
average of the individual values yielded $EW_{\rm B}/EW_{\rm
A}=0.73\pm0.05$ at a mean wavelength of $\lambda=5900$~\AA. Since the
effective temperatures of both components are very similar, as shown
below, the equivalent width ratio is analogous to luminosity ratio. From
this we derive values of $[L_{\rm B}/L_{\rm A}]_{\rm V}=0.72\pm0.05$,
$[L_{\rm B}/L_{\rm A}]_{\rm R}=0.74\pm0.05$ and $[L_{\rm B}/L_{\rm
A}]_{\rm I}=0.76\pm0.05$, for the luminosity ratios at the $V$, $R$ and
$I$ passbands, respectively. Our value for the $V$-band luminosity
ratio is indeed in good agreement with that obtained by D99. Although not
explicitly given in the paper, Delfosse (2000, priv. comm.) kindly
computed the luminosity ratio from their spectra and obtained a value of
0.68, which lies within one sigma of our determination.

In practice, the luminosity ratio or the ratio of radii is not one of the
parameters considered explicitly in the W-D code. Instead, are the surface
potentials of the stars (which, along with $q$, control the sizes) that
can be fixed or left free. We thus ran a number of solutions by adopting
different values for the potential of the primary star until reaching
agreement between the resulting luminosity ratio and the observed one.
This occurred for $\Omega_A=19.0$. All further solutions were ran by
fixing $\Omega_A$ at this value.

The shape of the out-of-eclipse region in the light curve clearly
indicates that one or both components have surface inhomogeneities. A
feature incorporated by the W-D program is the capability of fitting a
simple spot model. The spots, assumed to be circular in shape, are
described by W-D through four parameters, namely, the longitude, latitude,
angular radius, and temperature ratio (spot relative to photosphere). Some
of the spot parameters must be fixed when running solutions to prevent
lack of convergence. For example, the latitude of the spots can be
inferred theoretically. Granzer et al. (\cite{GSC00}) find that a star
such as the components of CU Cnc (0.4-M$_{\sun}$ star in the ZAMS with an
angular velocity 10 times the solar value) favours strongly the formation
of spots at latitudes of about 60\degr. Tests ran with W-D indeed
indicated a marginally better fit for spots at high latitudes.

As it seems obvious, the radius of the spot and the temperature contrast
are strongly correlated. No empirical information is available on the
temperature ratio between the spots and the surrounding photosphere for CU
Cnc. However, Hatzes (\cite{H95}) analysed Doppler tomography of the
M-type eclipsing binary YY Gem and found the presence of dark areas cooler
than the photosphere with $T_{\rm spot}/T_{\rm phot}=0.87-0.92$. For CU
Cnc, we expect similar values. We thus ran a series of solutions and
tested a variety of spot combinations (number of spots, locations, sizes,
temperature ratios, etc) to achieve the best possible fit to the light
curves (see a detailed discussion of a similar procedure in TR02). Our
preferred solution calls for two spots -- one on each component --.  The
spot on the primary component, which transits the central meridian at
orbital phase 0.438, appears to be relatively small, with a radius of
9\degr, and $\sim$450 K cooler than the photosphere. In contrast, the
secondary component has a much larger spot with a radius of 31\degr\ and a
temperature difference with the surrounding photosphere of $\sim$200 K.
The center of the dark area transits the central meridian at orbital phase
0.997. Its large size and relatively small temperature difference seems to
indicate that this is more likely a spot complex rather than a homogeneous
spot. Thus, the parameters determined represent an average over an
extended photospheric area probably covered with patchy dark spots.

\begin{figure}
\centering
\includegraphics[width=\linewidth]{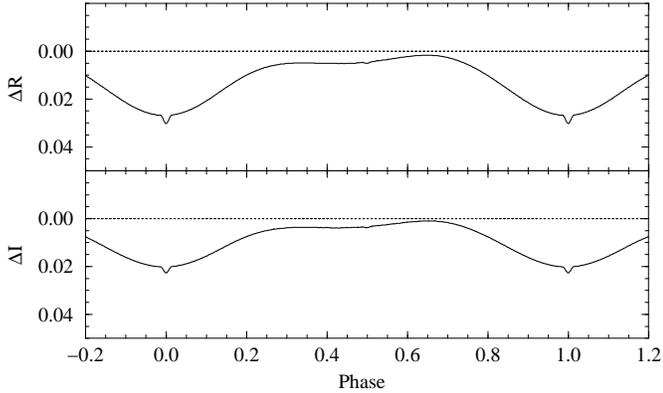}
\caption{Photometric effects of the spots on the $R$ and $I$ light curves. The
spike during the phase of the primary minimum arises when the primary component
is partially eclipsed and the spot on the secondary component is in front thus
enhancing its relative contribution.}
\label{figSpo}
\end{figure}

The photometric effect of the spot configuration adopted here is
illustrated in Fig. \ref{figSpo}. Note that the spot solution has only a
mild effect on the rest of the light curve parameters.  Furthermore, the
spot solution is not unique and other configurations may be equally valid
from a strictly numerical point of view.

\begin{figure}
\centering
\includegraphics[width=\linewidth]{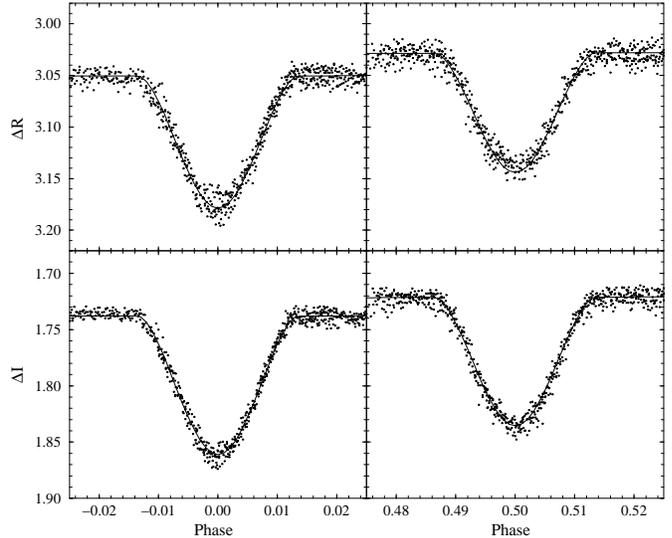}
\caption{Details of the fits to the primary and secondary eclipses. 
Note the high density of observations and the excellent agreement with the
synthetic light curve.}
\label{figEcl}
\end{figure}

The best-fitting two-spot model described above yields rms residuals of
0.008~mag and 0.005~mag for the $R$ and $I$ light curves, respectively.
These small values reflect both the quality of the photometry and the
excellent performance of the spot model. The synthetic light curve
superimposed to the observations can be seen in Fig. \ref{figLC}, together
with the residuals of the fit for each light curve. Also, for illustrative
purposes, we show a detail of the eclipse phases in Fig. \ref{figEcl}. The
solution indicates that the eclipses are partial and that $\sim$25\% of
each component's flux is blocked during the corresponding eclipses. The
parameters resulting from the light curve fit are listed in
Table~\ref{tabPars}. The uncertainties given in this table were not
adopted from the formal probable errors provided by the W-D code, but
instead from numerical simulations and other considerations. Several sets
of starting parameters were tried in order to explore the full extent of
the parameter space. In addition, the W-D iterations were not stopped
after a solution was found, instead, the program was kept running to test
the stability of the solution and the geometry of the $\chi^2$ function
near the minimum. The scatter in the resulting parameters from numerous
additional solutions yielded estimated uncertainties that we consider to
be more realistic, and are generally several times larger than the
internal statistical errors. For CU Cnc, however, an additional error
source must be considered and this is the uncertainty in the measurement
of the luminosity ratio. The empirical value obtained from the spectra
carries a small but significant measurement error that must be properly
accounted for. We did this by running new W-D solutions for values of
$\Omega_A$ set to yield the observed luminosity ratio plus and minus 1
$\sigma$. The average differences between the adopted and the new
parameters were combined quadratically with the uncertainties described
above to yield the values listed in Table~\ref{tabPars}. As a consequence
of the uncertainty in the luminosity ratio, the radii of the components
have larger errors ($\sim$2\%) than would be expected from the quality of
the light curve

\begin{table}
\centering
\caption[]{Results from the light curve analysis for CU Cnc.}
\label{tabPars}
\begin{tabular}{ll}
\hline
\multicolumn{1}{c}{Parameter}&\multicolumn{1}{c}{Value}\\
\hline
\multicolumn{2}{l}{Geometric and radiative parameters}\\
~~~$P$ (days) (fixed)                       & 2.771468            \\
~~~$e$ (fixed)                              & 0                   \\
~~~$i$ (deg)                                & 86.34$\pm$0.03      \\
~~~$q\equiv M_{\rm B}/M_{\rm A}$ (fixed)    & 0.9184              \\
~~~$\Omega_{\rm A}$ (fixed)                 & 19.00$\pm$0.22      \\
~~~$\Omega_{\rm B}$                         & 19.39$\pm$0.40      \\
~~~$r_{\rm A}$ (volume)                     & 0.0553$\pm$0.0007   \\
~~~$r_{\rm B}$ (volume)                     & 0.0501$\pm$0.0012   \\
~~~$k\equiv r_{\rm B}/r_{\rm A}$            & 0.906$\pm$0.025     \\
~~~$T_{\rm eff}^{\rm A}$ (K) (fixed)        & 3160                \\
~~~$T_{\rm eff}^{\rm B}/T_{\rm eff}^{\rm A}$& 0.9892$\pm$0.0019   \\
~~~$L_{\rm B}/L_{\rm A}$ ($R$ band, phase 0.287)& 0.74$\pm$0.05   \\
~~~$L_{\rm B}/L_{\rm A}$ ($I$ band, phase 0.287)& 0.76$\pm$0.05   \\
~~~$F_3$ ($R$ band, phase 0.215) (fixed)    & 0.23                \\
~~~$F_3$ ($I$ band, phase 0.215) (fixed)    & 0.25                \\
~~~Albedo (fixed)                           & 0.5                 \\
\vspace{2mm}
~~~Gravity brightening (fixed)              & 0.2                 \\
\multicolumn{2}{l}{Limb darkening coefficients (Logarithmic law)} \\
~~~$x_{\rm A}$ and $y_{\rm A}$ ($R$ band)   & 0.875~,~0.390       \\
~~~$x_{\rm B}$ and $y_{\rm B}$ ($R$ band)   & 0.875~,~0.390       \\
~~~$x_{\rm A}$ and $y_{\rm A}$ ($I$ band)   & 0.844~,~0.530       \\
\vspace{2mm}
~~~$x_{\rm B}$ and $y_{\rm B}$ ($I$ band)   & 0.844~,~0.530       \\
\multicolumn{2}{l}{Spot parameters}                               \\
~~~Phase for spot \#1                     &  0.438                \\
~~~Radius for spot \#1 (deg)              &  9                    \\
~~~$\Delta T$ for spot \#1 (K)            &  450                  \\
\vspace{1mm}
~~~Location of spot \#1                   &  Primary              \\
~~~Phase for spot \#2                     &  0.997                \\
~~~Radius for spot \#2 (deg)              &  31                   \\
~~~$\Delta T$ for spot \#2 (K)            &  200                  \\
\vspace{2mm}
~~~Location of spot \#2                   &  Secondary            \\
\multicolumn{2}{l}{r.m.s. residuals from the fits}                \\
~~~$\sigma_R$ (mag)                       & 0.008                 \\
~~~$\sigma_I$ (mag)                       & 0.005                 \\
\hline
\end{tabular}
\end{table}

The only independent determination of the light curve parameters of CU Cnc
comes from D99. As mentioned earlier, the authors only had sparse
observations and the quality of the solution is not very high.
Importantly, the effect of star spots were taken into account in the
analysis. The final radii published by D99 have errors of 12--16\%. Even
so, the agreement between D99's solution and the parameters listed in
Table \ref{tabPars} is mostly within their large quoted uncertainties. The
most significant difference is in the relative radii of the components,
where D99 report very unequal components ($k=0.67$). Recall that the ratio
of radii cannot be determined from the light curve alone so it is not
surprising that D99 found rather inconsistent values. The sum of the
relative radii, however, is tightly constrained by the duration of the
eclipses and our value (0.1054$\pm$0.0006) is extremely close to that
found by D99 (0.1052).

\section{Absolute dimensions and effective temperatures} \label{secAbs}

The absolute dimensions of the CU Cnc components can be readily determined
by combining the photometric parameters in Table \ref{tabPars} with the
spectroscopic solution from D99. When doing so, we obtain the physical
properties listed in Table \ref{tabProps}.

\begin{table}
\centering
\caption[]{Absolute dimensions and radiative properties for the components of
CU Cnc.}
\label{tabProps}
\begin{tabular}{lll}
\hline
\multicolumn{1}{c}{Property}&\multicolumn{1}{c}{Primary}&\multicolumn{1}{c}{Secondary}\\
\hline
Spectral type        &        M3.5 Ve           &          M3.5 Ve         \\
Mass (M$_{\sun}$)    &        0.4333$\pm$0.0017 &        0.3980$\pm$0.0014 \\
Radius (R$_{\sun}$)  &        0.4317$\pm$0.0052 &        0.3908$\pm$0.0094 \\
$\log g$ (cgs)       &        4.804$\pm$0.011   &        4.854$\pm$0.021   \\
T$_{\rm eff}$ (K)    &        3160$\pm$150      &        3125$\pm$150      \\
$\log (L/L_{\sun})$  & \llap{$-$}1.778$\pm$0.083& \llap{$-$}1.884$\pm$0.086\\
$M_{\rm bol}$ (mag)  &        9.19$\pm$0.21     &        9.45$\pm$0.21     \\
$\pi$ (mas)          &     \multicolumn{2}{c}{78.05$\pm$5.69}              \\
$M_{\rm V}$ (mag)    &        11.95$\pm$0.16    &        12.31$\pm$0.16    \\
$BC_{\rm V}$ (mag)   & \llap{$-$}2.76$\pm$0.26  & \llap{$-$}2.86$\pm$0.26  \\
\hline
\end{tabular}
\end{table}

The absolute value of the effective temperature for eclipsing
binaries cannot be determined from the light curves alone and independent
methods must be used. Temperature determination is usually quite
problematic when one has to deal with very cool stars such as CU Cnc. Two
independent approaches present themselves to estimate the effective
temperature of CU Cnc: one based on photometric calibrations and one based
on the known absolute dimensions and distance.

Synthetic colors, which are the base of photometric calibrations, are
computed by modelling the stellar atmospheres. Albeit some discrepancies
remain, M-star model atmospheres have made significant progress over the
past decade (Allard et al. \cite{AHA97}). Line opacities from a large
number of molecular species are now included in the calculations and
substantial improvements have been made in the input physics, particularly
the equation of state. For cool stars, the infrared colors are especially
suited to estimate the effective temperature because most of the stellar
flux is emitted in that wavelength range. Thus, we collected from the
literature the available magnitudes and colors for CU Cnc, which are
listed in Table \ref{tabCol}. Simple linear transformations were applied
to refer the observed magnitudes and colors to the photometric systems
most widely used.

\begin{table*}
\centering
\caption[]{Joint observed magnitudes and color indices for CU Cnc.}
\label{tabCol}
\begin{tabular}{lll}
\hline
\multicolumn{1}{c}{Mag. or color}&\multicolumn{1}{c}{Value}&
\multicolumn{1}{c}{Notes}\\
\hline
$V$                           &$\llap{1}1.89\pm0.01$& From Weis (\cite{W91}) \\
$(R-I)_{\rm C}^{\mathrm{a}}$  &  $1.60$             & From Weis (\cite{W91}) and transformations in Bessell (\cite{B79}) \\
$(V-I)_{\rm C}^{\mathrm{a}}$  &  $2.80$             & From Weis (\cite{W91}) and transformations in Bessell (\cite{B79}) \\
$J_{\rm 2MASS}$               & $7.536\pm0.018$     & 2MASS Second Incremental Data Release$^{\mathrm{b}}$ \\
$H_{\rm 2MASS}$               & $6.912\pm0.030$     & 2MASS Second Incremental Data Release$^{\mathrm{b}}$ \\
$K_{\rm 2MASS}$               & $6.626\pm0.029$     & 2MASS Second Incremental Data Release$^{\mathrm{b}}$ \\
$K_{\rm BB}^{\mathrm{c}}$     & $6.670\pm0.031$     & Using transformations from Carpenter (\cite{C01}) \\
$K_{\rm CIT}^{\mathrm{d}}$    & $6.650\pm0.031$     & Using transformations from Carpenter (\cite{C01}) \\
$(V-K)_{\rm BB}^{\mathrm{c}}$ & $5.22\pm0.03$       &                                            \\
$(V-K)_{\rm CIT}^{\mathrm{d}}$& $5.24\pm0.03$       &                                            \\
\hline
\end{tabular}
\begin{list}{}{}
\item[$^{\mathrm{a}}$] Cousins (\cite{C76}) photometric system
\item[$^{\mathrm{b}}$] {\tt http://www.ipac.caltech.edu/2mass/releases/second/doc/explsup.html}
\item[$^{\mathrm{c}}$] Bessell \& Brett (\cite{BB88}) homogenized system
\item[$^{\mathrm{d}}$] Caltech photometric system (Elias et al. \cite{EFM82}, \cite{EFH83})
\end{list}
\end{table*}

The magnitudes and color indices in Table \ref{tabCol} correspond to the
combined light of the two components of CU Cnc. Since the effective
temperatures of the two components are very similar, it is justified to
use the joint color indices as those representing the ``mean'' component
of CU Cnc. To estimate the mean effective temperature we employed a number
of modern photometric calibrations.  We assumed no interstellar reddening
in our calculations as expected for a system that lies less than 20 pc
away. The resulting temperatures, together with the references, atmosphere
model names, and comments, are presented in Table \ref{tabTeff}. Most of
the calibrations are based upon synthetic colors and model atmosphere
calculations, except for Leggett (\cite{L92}), who provides an empirical
calibration. There is remarkably good agreement between all the
temperature estimates. A plain average of the independent values (i.e.,
excluding one of the redundant temperatures from the BaSeL model) yields
$T_{\rm eff}=3140\pm40$ K. The formal error, however, does not reflect any
possible systematics that could be present in the calibration. We have
thus adopted a more realistic error value of $T_{\rm eff}=3140\pm150$ K
based on the results of Leggett et al. (\cite{LAB96}).

\begin{table*}
\centering
\caption[]{Effective temperature determinations for the mean component of
CU Cnc.}
\label{tabTeff}
\begin{tabular}{lccl}
\hline
\multicolumn{1}{c}{Reference}&\multicolumn{1}{c}{Model}&
\multicolumn{1}{c}{$T_{\rm eff}$ (K)}&\multicolumn{1}{c}{Colors used}\\
\hline
Allard et al. (\cite{AHS00})     &  STARDusty2000 &  3130   & $(V-I)_{\rm C}$, $(V-K)_{\rm CIT}$ \\
Hauschildt et al. (\cite{HAB99}) &  NextGen       &  3170   & $(V-K)_{\rm CIT}$ \\
Lejeune et al. (\cite{LCB98})    &  BaSeL 3.1     &  3170   & $(V-I)_{\rm C}$, $(V-K)_{\rm BB}$ \\
Lejeune et al. (\cite{LCB98})    &  BaSeL 2.2     &  3170   & $(V-I)_{\rm C}$, $(V-K)_{\rm BB}$ \\
Bessell et al. (\cite{BCP98})    &  NMARCS        &  3070   & $(V-I)_{\rm C}$, $(V-K)_{\rm BB}$ \\
Leggett (\cite{L92})             &  --            &  3145   & $(R-I)_{\rm C}$, $(V-I)_{\rm C}$, $(V-K)_{\rm CIT}$ \\
\hline
\end{tabular}
\end{table*}

As mentioned above, there is an alternative approach to compute the
temperature of the components based on the method described in Ribas et
al. (\cite{RGT98}). This is a simple method that relates the observed
radius and luminosity with the temperature. In the case of CU Cnc, the
absolute radii of the components are given in Table \ref{tabProps}. The
empirical luminosities were computed from the apparent magnitudes, the
parallax in the Hipparcos catalogue (ESA \cite{ESA97}) and a bolometric
correction. The latter is especially delicate because bolometric
correction calibrations are known to be the source of systematic errors.
To minimise these, we used the bolometric correction in the $K$ band,
which was computed from the models of Allard et al. (\cite{AHS00}) and
resulted in ${BC_{\rm K}}_{\rm BB}=+2.75\pm0.08$ mag. This leads to an
apparent bolometric luminosity of $m_{\rm bol}=9.42\pm0.08$. From the
luminosity ratio, the individual bolometric magnitudes can be computed and
we find ${m_{\rm bol}}_{\rm A}=10.04\pm0.08$ and ${m_{\rm bol}}_{\rm
B}=10.31\pm0.08$. The absolute bolometric magnitudes follow from
introducing the Hipparcos distance: ${M_{\rm bol}}_{\rm A}=9.50\pm0.18$
and ${M_{\rm bol}}_{\rm B}=9.77\pm0.18$. The effective temperature of each
component can be computed in a straightforward manner and we obtain
${T_{\rm eff}}_{\rm A}=2950\pm120$ K and ${T_{\rm eff}}_{\rm
B}=2910\pm120$ K.

The agreement with the photometric determinations described above is not
perfect but nevertheless fairly good given the uncertainties. The values
agree within one sigma of their respective error bars. However, a caveat
is due at this point. As it turns out, there are reasonable concerns about
the accuracy of the Hipparcos parallax, even taking into account its
relatively large error (see Table \ref{tabProps}). CU Cnc is a faint
object near the threshold for detection by Hipparcos. Thus, the individual
transit data are of poor quality. A reanalysis of the transit data
(considering also the visual component) has not provided any better
results (Arenou 1999, priv. comm.).  As a reference, we have computed the
temperature of CU Cnc as a function of the distance\footnote{Note that
there is a weak dependence of the bolometric correction on the
temperature, so these expressions are only valid for temperatures not too
far off from those in Table \ref{tabProps}.}:
\[{T_{\rm eff}}_{\rm A}=2610 \sqrt{\frac{d\mbox{(pc)}}{10}} \;\;\mbox{K;}
\hspace{1cm}
{T_{\rm eff}}_{\rm B}=2575 \sqrt{\frac{d\mbox{(pc)}}{10}} \;\;\mbox{K}\] 

Given the concerns with the Hipparcos distance raised above, we prefer to
adopt the temperatures derived from photometric calibrations. The final
adopted individual temperatures for CU Cnc (computed from the average
temperature and the temperature ratio derived from the light curves) are
listed in Table \ref{tabProps} along with their associated errors. Also
included in Table \ref{tabProps} are the absolute magnitudes of the
components in the $V$ band, which have been computed from the parallax
listed in the Hipparcos catalogue. Note that an empirical determination of
the bolometric correction for each component follows from the comparison
of the $V$-band absolute magnitude with the bolometric magnitudes
(resulting from the radii, effective temperatures and ${M_{\rm
bol}}_{\sun}=4.74$).

\section{Other observational information}
\label{secAge}

\subsection{Age and chemical composition}

One of the goals of the present study is to carry out a critical analysis
of low-mass stellar models similar to that in TR02. Thus, to make the test
as stringent as possible, one has to keep the number of degrees of freedom
at a minimum. Two of the key parameters that determine the observable
physical properties of a star are its age and chemical composition (to
simplify, metallicity). The metallicity of a star is generally determined
through high-resolution spectroscopy and comparison with synthetic
spectra. Unfortunately, this is virtually impossible for M stars because
of the huge quantity of spectral features and the shortcomings of the
atmosphere models. In turn, the age (and eventually the metallicity) can
only be derived if the star belongs in a well-studied cluster. Field M
stars are, thus, not ideally suited for reliable estimations of age and
metallicity.

One possible way to obtain a rough estimation of the age and chemical
composition of a field M star such as CU Cnc is through the analysis of
its space velocities. High velocity would likely indicate an (old) halo
population and low velocity a (young) star in the galactic disk. The space
velocities $(U,V,W)$ of a star are readily computed from its position,
proper motions, radial velocity and distance. In the case of CU Cnc, we
have adopted the Hipparcos position, proper motion and distance, and the
radial velocity has been taken from D99. The resulting space velocities
are $U=-10.7$ km~s$^{-1}$, $V=-4.9$ km~s$^{-1}$, and $W=-10.6$
km~s$^{-1}$. Note that we have followed the convention where positive
values of $U$, $V$, and $W$ indicate velocities towards the galactic
center, galactic rotation and galactic North Pole, respectively.

The low values of these velocities are an indication of a disk population
and thus a relatively young star. A closer look at the values revealed
that these are strikingly similar to the velocities of the Castor moving
group (see Anosova \& Orlov \cite{AO91}; Barrado y Navascu\'es \cite{B98})
of which YY Gem is a likely member. Ribas \& Rebolo (\cite{RR02}) utilised
various criteria (isochrones, rotation-activity relationships, etc) to
estimate an age of 320$\pm$80 Myr and a metal content of $Z=Z_{\odot}$ for
the moving group members. The mean space velocities of the moving group
have been recently re-determined by Ribas \& Rebolo (\cite{RR02}) to be
$<U>=-10.6\pm3.7$ km~s$^{-1}$, $<V>=-6.8\pm2.3$ km~s$^{-1}$, and
$<W>=-9.4\pm2.1$ km~s$^{-1}$. These are indeed very similar to the values
found for CU Cnc, which strongly suggests its membership of the Castor
moving group. Thus, it is sensible to assume that CU Cnc has the same age
and chemical composition as the rest of the group members -- similarly to
what is routinely done in stellar cluster studies. As it happened with YY
Gem (TR02) the fortunate fact that CU Cnc belongs in the moving group and
has a well-determined age and metal content greatly enhances the value of
the star since it allows for an unusually stringent test of the
theoretical models.

\subsection{Lithium abundance}

As it has been known for some time, the measured abundance of lithium in
the stellar atmosphere can be used as an age indicator for young cool
stars, particularly in clusters (e.g., Duncan \cite{D81}; Soderblom et al.
\cite{SOJ90}). Theoretical studies demonstrate that the $^7$Li nuclei are
destroyed by proton collisions at temperatures above $\sim$2.5 MK
(Bodenheimer \cite{B65}; D'Antona \& Mazzitelli \cite{DM84}). In low mass
stars, mixing through convection processes is so efficient that Li is
depleted at a very rapid pace. We have inspected our high-resolution
spectrum described in Sect. \ref{secSol} and we report a tentative
detection of weak Li~{\sc i} $\lambda$6708 features at the expected
wavelengths for both the primary and secondary components. The equivalent
width measurement (corrected for light dilution) yields a value of
$\sim$50 m\AA\ for both stars. From this and the theoretical study of
Pavlenko et al. (\cite{PRM95}), one can roughly estimate the Li abundance
to be $\log N(\mbox{Li})\approx -1.1$ (in the scale in which $\log
N(\mbox{H})=12$), and $\log [n({\rm Li})/n_{\circ}({\rm Li})]=-4.1$. Thus,
the initial Li content of CU Cnc's components appears to have been mostly
destroyed, yet not completely. This is in apparent contradiction with Li
destruction sequences in clusters and associations (Barrado y Navascu\'es
et al. \cite{BSP99}; Stauffer et al. \cite{SBB99}), which indicate that
mid-M type stars fully deplete their initial Li abundance in as little as
a few times $10^7$ years. (Recall that the estimated age for CU Cnc is
$\sim$320 Myr.) Barrado y Navascu\'es et al. (\cite{BFG97}) observed
a similar discrepancy with the observed Li abundance ($\log
N(\mbox{Li})=0.0$) for the eclipsing binary YY Gem. In this case, Li also
appears to be severely depleted but not completely destroyed, as would be
expected for YY Gem's age of $\sim$320 Myr. If we rely on the ages
determined for these binaries, the two results seem to indicate that Li
depletion for M stars in binary systems might be not as efficient as for
single stars.  For example, orbital synchronization due to tidal forces
could be responsible for inhibiting turbulent mixing, thus slowing down Li
depletion (Barrado y Navascu\'es et al. \cite{BFG97}). The tentative
detection of Li in the spectrum of CU Cnc is a puzzle yet to be resolved a
detailed discussion of this topic is left for an upcoming work.

\subsection{Stellar activity}

Other observational information for CU Cnc can be inferred from the
high-resolution spectrum described in Sect. \ref{secSol}. An inspection of
the spectrum reveals very strong H$\alpha$ and H$\beta$ emission features
(double lines). The measurements yield (continuum-corrected) H$\alpha$
equivalent widths of 3.85~\AA\ and 4.05~\AA\ for the primary and secondary
components, respectively. For comparison, Young et al. (\cite{YSS89})
report an excess equivalent width of $\approx$2~\AA\ for the presumably
coeval but more massive system YY Gem. The H$\alpha$ equivalent widths of
both CU Cnc and YY Gem are rather large but not unreasonable when compared
with young M-type stars (see, e.g., Soderblom et al. \cite{SDJ91}). Note,
however, that the strong H$\alpha$ emission in these binary systems is
more related to the spin-up caused by orbital synchronization rather than
age.

In addition to displaying H$\alpha$ chromospheric emission, CU Cnc is also
both and EUV and X-ray source. From the observations in the ROSAT All-Sky
Bright Source Catalogue (Voges et al. \cite{Vea99}) and the calibration in
Schmitt et al. (\cite{SFG95}) one obtains an integrated X-ray luminosity
of $\log L_{\rm X} \mbox{(erg s$^{-1}$})=29.03$ for CU Cnc.  Again, this
value can be compared with YY Gem's, which was estimated to be $\log
L_{\rm X} \mbox{(erg s$^{-1}$})=29.61$ (G\"udel et al. \cite{GSB93}). One
must keep in mind that the size of the stars in CU Cnc is significantly
smaller than those of YY Gem.  Thus, the ratio $L_{\rm X}/L_{\rm bol}$ is
a more realistic measure of the stellar activity or, in other words, of
the efficiency in producing high energy emissions. The calculations of
$L_{\rm X}/L_{\rm bol}$ yield a value very close to $10^{-3}$ for both CU
Cnc and YY Gem, thus indicating very similar activity levels. Numerous
studies (see, e.g., Jeffries \& Tolley \cite{JT98}; Gagne et al.
\cite{GCS95}) suggest that this value of $L_{\rm X}/L_{\rm bol}=10^{-3}$
indeed marks a saturation limit in the stellar activity.

\section{Critical comparison with stellar models}

CU Cnc is only the third M-type eclipsing binary for which accurate
absolute dimensions (masses {\em and} radii) have been determined. In
addition, we have been able to set rather stringent constraints on the
effective temperatures of the components as well as on the age and
chemical composition of the system. With this information in hand, we are
in a position to carry out a critical evaluation of the stellar model
predictions. For the first time we can provide observational checks on M
stars of about 0.4~M$_{\sun}$ for which no reliable empirical information
has been available to date. The procedure followed here is very similar to
that described in TR02 for YY~Gem, but with a significant difference.
While the two components of YY Gem are almost identical, this is not the
case for CU Cnc where the components differ by almost 10\% in mass. Since
the two stars in the binary system are coeval, a single isochrone expected
to fit both components simultaneously. This implies that not only the
position in the observational diagrams but also the slopes of the
isochrones come into play. Thus, the models must reproduce both the
absolute and the relative location of the components.

For our comparison we have considered nine different sets of theoretical
calculations: Swenson et al. (\cite{SFR94}), D'Antona \& Mazzitelli
(\cite{DM97}), Siess et al. (\cite{SFD97}), Baraffe et al. (\cite{BCA98}),
Palla \& Stahler (\cite{PS99}), Charbonnel et al. (\cite{CDS99}), Girardi
et al. (\cite{GBB00}), Yi et al. (\cite{YDK01}), and Bergbusch \&
VandenBerg (\cite{BV01}). These include essentially all of the modern
models for low mass stars, even though some of them do not quite reach
down to the masses of CU Cnc. We have, nevertheless, incorporated them so
that their performance at higher masses can be compared with the rest of
the models. Isochrones for an age of 320~Myr and $Z = 0.018$ have been
interpolated for all models except those by Palla \& Stahler
(\cite{PS99}), with an oldest age available of 100~Myr (this has no effect
whatsoever on our conclusions). The models of Siess et al. (\cite{SFD97})
and Baraffe et al. (\cite{BCA98}) employ the most sophisticated physics
(equation of state) and boundary conditions (non-grey model atmospheres),
so, in principle, they would be expected to provide the best fit to the
physical properties of CU Cnc.

   \begin{figure*}
   \centering
   \includegraphics[width=7.3cm]{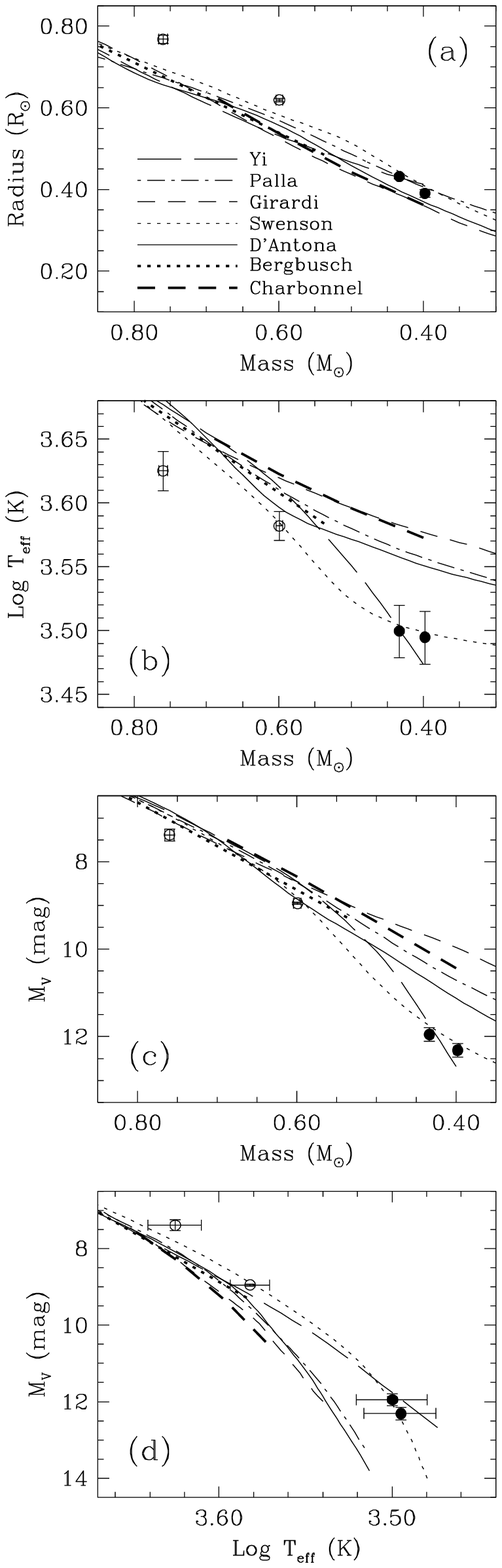}
\hspace*{1cm}
   \includegraphics[width=7.3cm]{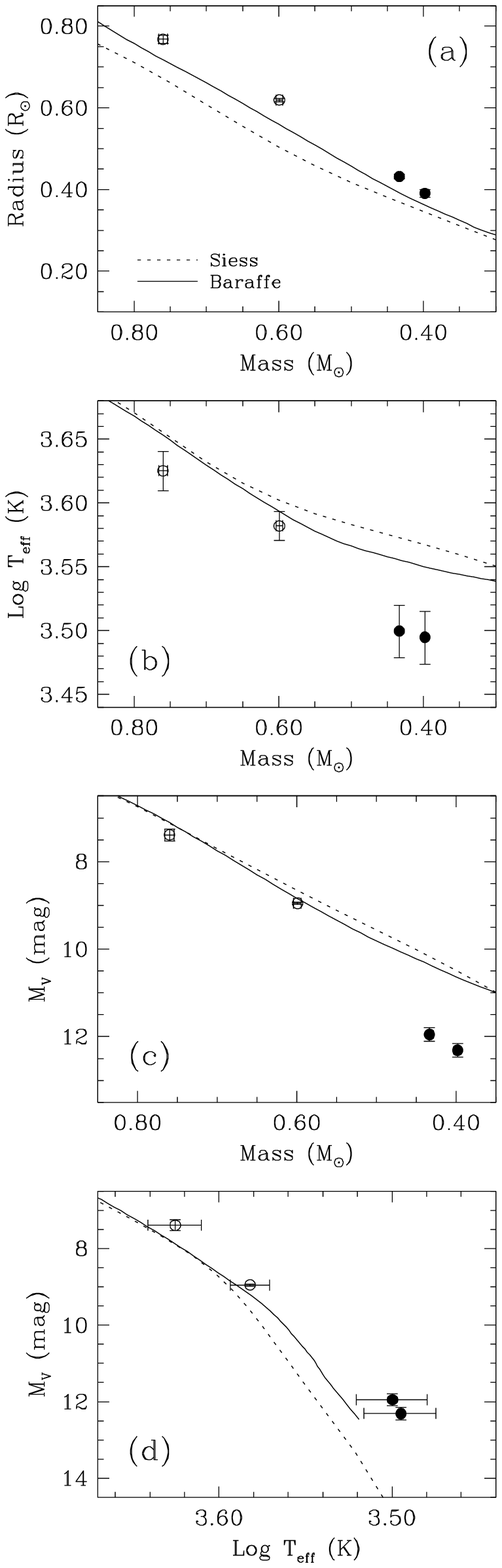}
      \caption{The location of CU Cnc's components on several observational
planes is compared with isochrones from nine different theoretical models (see
legends and references in text). The age for the isochrones is 320~Myr and the
metallicity adopted is $Z=0.018$, except for the isochrones by Palla \& Stahler
(\cite{PS99}) that are for an age of 100~Myr and solar abundance. In addition
to the components of CU Cnc (represented with filled circles), we show the
locations of YY Gem's mean component at a mass of 0.6~M$_{\sun}$ and the
secondary component of V818 Tau with a mass of 0.76~M$_{\sun}$ (both
represented as open circles).}
         \label{figMod}
   \end{figure*}

We have compared the observational data for CU Cnc with the model
predictions at different planes. The plots are shown in the eight panels
of Fig. \ref{figMod}. For clarity reasons, we have plotted the models of
Siess et al. (\cite{SFD97}) and Baraffe et al. (\cite{BCA98}) separated
from the rest in the right panels. Note that we have also included the
observational data on YY~Gem~AB and V818~Tau~B (both from TR02), which
will help to understand the comparison in a wider context. The best
determined parameters of our analysis are the masses and the radii, which
are purely empirical, whereas the effective temperatures and absolute
magnitudes hinge to some extent on external calibrations. Thus, the
mass-radius diagram in Fig. \ref{figMod}a constitutes the most reliable
check on the models, also because the error bars of the measurements are
very small. As can be seen, some of the models yield reasonably good,
albeit not perfect, fits to the data. The three isochrones that exhibit
the best performance are those by Swenson et al. (\cite{SFR94}), Palla \&
Stahler (\cite{PS99}), and Yi et al. (\cite{YDK01}). However, among these,
only the models of Swenson et al. (\cite{SFR94}) succeed in reproducing at
the same time the {\em relative} location (i.e. slope) of the binary
components. The models of Siess et al. (\cite{SFD97}) and Baraffe et al.
(\cite{BCA98}), in spite of using state-of-the-art physical ingredients,
do not yield a good fit to the data, although the latter perform slightly
better. The radii that these models predict at the CU Cnc's masses are
about 10--14\% smaller than observed. (Note that this difference amounts
to $\sim$7$\sigma$.) The plots indicate that the huge discrepancies
between model predictions and observations encountered by TR02 at slightly
higher masses (YY~Gem~AB and V818~Tau~B) are not quite as dramatic at the
mass regime of CU Cnc, and some of the models produce reasonable fits to
the data.

The panels in Fig. \ref{figMod}b introduce the effective temperature in
the comparison. Here the internal discrepancies among the seven models
grow larger as we move towards lower masses. At around 0.4~M$_{\sun}$
differences of up to 20\% exist in the effective temperatures predicted by
the different isochrones. Interestingly, most of the models seem to
indicate temperatures that are significantly (10--15\%) higher than
observed for CU Cnc. The isochrones of Swenson et al. (\cite{SFR94}) and
Yi et al. (\cite{YDK01}) clearly stand out among all and appear to produce
a fairly close fit. But again, only the isochrone computed from the models
by Swenson et al. (\cite{SFR94}) reproduces the relative location of the
components to achieve an outstanding fit to CU Cnc and also YY Gem,
although V818 Tau~B seems significantly cooler than model predictions.

The mass-luminosity diagram in Fig. \ref{figMod}c presents a situation
which is analogous to that described above. Most of the isochrones predict
$M_{\rm V}$'s for CU Cnc that are too bright by up to 2 magnitudes. For
example, the models of Siess et al. (\cite{SFD97}) and Baraffe et al.
(\cite{BCA98}) yield close fits to both V818 Tau~B and YY Gem while huge
differences are found with the observed magnitudes of CU Cnc.  The models
by Swenson et al.  (\cite{SFR94}) are able to both reproduce the absolute
magnitude of CU Cnc {\em and} YY Gem, while also achieving a quite close
fit to V818 Tau~B.  At the same time, the slope of the isochrone matches
perfectly the relative configuration of the CU Cnc components.

Finally, the H-R diagrams in Fig. \ref{figMod}d do not reserve any
surprises. Most of the models tend to overestimate the effective
temperature at a given absolute magnitude. The best agreement is found for
the models of Swenson et al. (\cite{SFR94}) that produce an isochrone with
the correct slope that reproduces well the observed parameters of CU Cnc
and YY Gem. Note that the relatively close fit by the models of Baraffe et
al. (\cite{BCA98}) is fictitious because it occurs for masses that are far
off from those determined for CU Cnc.

\section{Is CU Cnc akin to other stars of similar mass?}

\begin{figure}
\centering
\includegraphics[width=\linewidth]{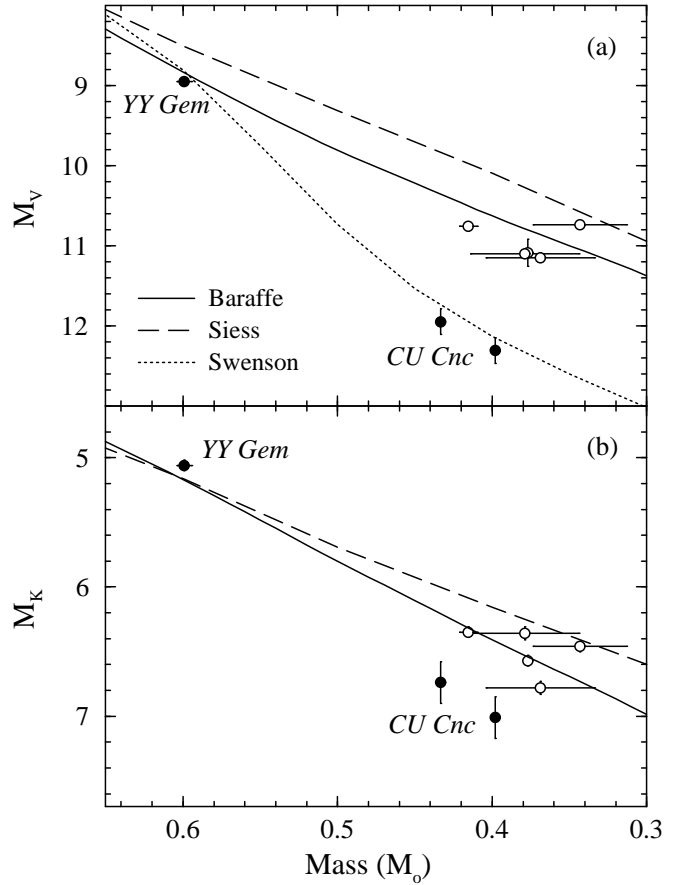}
\caption{Mass-absolute magnitude diagrams in the $V$ and $K$ bands for the
lower main sequence. CU Cnc is shown together with YY Gem and other stars with
spectroscopically-determined masses (see text). The lines represent isochrones
for the estimated age of CU Cnc computed with different models.}
\label{figMKM}
\end{figure}

The comparison discussed above leaves no doubt that most of current
stellar models fail in reproducing the observed physical properties of CU
Cnc. Only the models by Swenson et al. (\cite{SFR94}) yield reasonably
good (even excellent) isochrone fits to the data. Even though CU Cnc is
the only object with a mass around 0.4~M$_{\sun}$ with fully characterised
physical properties, there are other stars in this regime whose masses
have been spectroscopically determined to good accuracy. This is the case
of \object{Gl 570C}, \object{Gl 623A}, \object{Gl 644A}, \object{Gl
644Ba}, \object{Gl 661A}, and \object{Gl 661B}, all with masses between
0.34 and 0.42~M$_{\sun}$ (see Forveille et al. \cite{FBD99}; S\'egransan
et al.  \cite{SDF00}; Delfosse et al. \cite{DFS00}; Martin et al.
\cite{MHM98}; Mazeh et al. \cite{MLG01}). We have compiled $V$ and $K$
absolute magnitudes for these stars from Delfosse et al. (\cite{DFS00}).
Two mass--absolute magnitude plots are shown in Fig. \ref{figMKM}, one for
the $V$ magnitude and one for the $K$ magnitude.  The results are quite
striking as CU Cnc appears to be fainter in both plots than other stars of
the same mass. The magnitude difference is about 1.4 mag in the $V$ band
and 0.35 mag in the $K$ band. The apparent faintness of CU Cnc in the $V$
band had already been pointed out by D99, Delfosse et al.  (\cite{DFS00})
and Mazeh et al. (\cite{MLG01}). Note that the isochrones computed with
the models of Baraffe et al. (\cite{BCA98}) and Siess et al.  
(\cite{SFD97}) plotted in the figure produce a reasonably close fit to the
stars other than CU Cnc in both observational diagrams. On the other hand,
the isochrone by Swenson et al. (\cite{SFR94}) -- only available in
$M_{\rm V}$ --, which closely matches CU Cnc, fails in reproducing the
observed absolute magnitudes of the other stars.

Let us carefully analyse possible reasons for the discrepancy. One
scenario to explain the disagreement is the effect of enhanced stellar
activity in CU Cnc due to binarity (higher rotational velocity).  This
would cause the appearance of large spots on the surface of the stars and
also an increase of the TiO absorption. It is well known that the $V$
passband is strongly affected by TiO absorption whereas the $K$ band is
mostly free from TiO bands. This might cause the star to appear especially
faint in the $V$ band. There are, however, evidences that argue against
this scenario. All observational data (X-ray luminosity, H$\alpha$
emission) indicate that CU Cnc is not especially active when compared with
YY Gem, which is quite closely matched by the models. At CU Cnc's mass
range, the ROSAT All-Sky Bright Source Catalogue (Voges et al.
\cite{Vea99}) yields an estimated X-ray luminosity for the stars in the
\object{Gliese 644} system (plotted in Fig. \ref{figMKM}) of $\log L_{\rm
X} \mbox{(erg s$^{-1}$})=29.14$, thus indicating an activity level very
similar to CU Cnc's.  In addition, the available multi-band photometry
(see Table \ref{tabCol}) refutes the idea that the apparent faintness of
CU Cnc is only restricted to the $V$ band. The spectral energy
distribution deduced from the photometric data is internally consistent,
i.e. all photometric indices yield the same effective temperature. Thus,
it is the temperature of CU Cnc and not the photometry the source of the
discrepancy. Using $(V-K)$ indices and the calibrations in Sect.
\ref{secAbs}, we estimate that CU Cnc is some 300~K cooler than other
stars of about 0.4~M$_{\sun}$. That the mean photospheric temperature is
reduced by as much as 10\% due to enhanced stellar activity in CU Cnc
seems, in principle, quite unlikely. More so if we take into account that
stellar activity produces photospheric plages in addition to dark spots
that would (partly or totally) compensate for the temperature reduction.

The distance to CU Cnc is another possible source for discrepancies since,
as we have discussed in Sect. \ref{secAbs}, it might be subject to
systematic errors. However, a magnitude differential in the distance
modulus of the star would be constant over wavelength and affect
identically both $V$ and $K$ bands. This is a strong argument against this
scenario because we observe a much larger deviation in $M_{\rm V}$ than in
$M_{\rm K}$. Nonetheless, a problem with the distance could still be
responsible for a fraction of the discrepancy.

Our estimate for the metal abundance for CU Cnc comes from its likely
membership of the Castor moving group. The possibility for a different
chemical composition can be not completely discarded because the star
might not belong in the group after all or have different abundances than
other group members. A very high metal abundance of $[\mbox{Fe/H}]=+0.5$
was put forward by D99 as a possible explanation for the anomalous
location of CU Cnc in the mass--absolute magnitude diagram. This was
proposed on the basis of the isochrone behaviour in the mass-$M_{\rm V}$
diagram for metal abundances of $[\mbox{Fe/H}]=0$ and
$[\mbox{Fe/H}]=-0.5$. We have carried out this analysis again with the new
parameters for CU Cnc and the models of Baraffe et al. (\cite{BCA98}). If,
as assumed by D99, the absolute magnitude changes are supposed to scale
linearly with metallicity, a value even higher than $+0.5$ is necessary
since the $V$ magnitude difference between $[\mbox{Fe/H}]=0$ and
$[\mbox{Fe/H}]=-0.5$ at a mass of 0.4~M$_{\sun}$ is only 0.8 mag. We
recall that CU Cnc appears to be dimmer by as much as 1.4 mag.
Furthermore, this scenario does not explain the discrepancy in the $K$
band because the different metallicity isochrones run virtually on top of
one another. An additional argument against the high-metallicity
hypothesis is the lack of other stars in the solar neighbourhood with such
high metal abundance. In the thorough study of Edvardsson et al.
(\cite{EAG93}) the highest metallicity found was $[\mbox{Fe/H}]=+0.24$.
The catalog of Cayrel de Strobel et al. (\cite{CdS01}) lists only a
handful of stars with $[\mbox{Fe/H}]>+0.4$ and most of them appear to be
misclassified. The high metallicity proposed to resolve the issue with the
location of CU Cnc in the mass-$M_{\rm V}$ diagram seems, at this point,
quite unlikely.

A younger age than estimated for CU Cnc from its membership of the Castor
moving group would neither explain the discrepancy. A low mass star
becomes dimmer as it evolves from the pre-main sequence phase towards the
ZAMS, where the minimum luminosity is reached. CU Cnc is already fainter
than the other stars in the same mass range. Alternatively, it could be
that the stars used for comparison with CU Cnc (Gl 570, Gl 623, Gl 644, Gl
661) were all much younger ($\sim$10--20 Myr) and contracting towards the
ZAMS. Once again, this hypothetical situation is not in agreement with the
observed mass-magnitude diagrams because the magnitude change in the $V$
and $K$ bands are predicted by the models to be very similar (actually, a
somewhat larger magnitude variation in the $K$ band).

Thus far, there are no plausible scenarios to account for the observed
magnitude differences between CU Cnc and similar mass stars. Even though
it might appear as improbable, the presence of a certain amount of dust
absorption could reproduce the observed behaviour. Indeed, according to
Fitzpatrick (\cite{F99}), the relationship between $A_{\rm V}$ and $A_{\rm
K}$ for interstellar dust is $A_{\rm V}\approx 8\; A_{\rm K}$. Thus, 1.4
magnitudes of extinction in $V$ would roughly correspond to 0.2 magnitudes
in $K$. Such magnitude corrections would bring the two mass-magnitude
plots into fair agreement. At a distance of only $\sim$13 pc, it can be
definitely ruled out that line-of-sight interstellar dust could account
for over 1 magnitude of extinction in the $V$ band ($E(B-V)=0.42)$. The
alternative is therefore circumstellar dust. Disks of circumstellar dust
have been directly detected in a number of nearby stars (see, e.g.,
Greaves et al. \cite{GHM98} and references therein) and some of them have
estimated ages similar to CU Cnc's. If the disk was coplanar with the
binary's orbit, as it might likely be assumed, we could be observing the
stars through a line of sight with enhanced dust absorption. There are no
observational grounds to support such hypothesis yet but, at this point,
this is the only scenario that explains the observations.

If the disk hypothesis is proven to be correct, our temperature estimate
(which is based on multi-colour photometry) and thus the comparison with
stellar models in the temperature and absolute magnitude diagrams would
have to be revised accordingly. A rough estimate made by correcting for
the differences of 1.4 mag in $V$ and 0.35 mag in $K$ raises the
temperature of the components to ${T_{\rm eff}}_{\rm A}=3440\pm150$ K and
${T_{\rm eff}}_{\rm B}=3400\pm150$ K. Also, the effective temperatures
obtained from the radii, the corrected apparent magnitudes and the
Hipparcos parallax (see Sect. \ref{secAbs}) are ${T_{\rm eff}}_{\rm
A}=3290\pm140$ K and ${T_{\rm eff}}_{\rm B}=3240\pm140$ K, thus achieving
a remarkable 1 sigma agreement with the photometric values.

\section{Conclusions}

The lack of observational constraints is one of the most severe
deficiencies that stellar models for the lower main sequence have to face.
Eclipsing binaries are thus crucial because they yield simultaneous
determinations of masses and radii (and also temperatures) upon which the
stellar models can be checked. We have therefore acquired and analysed
high-quality $R$ and $I$ light curves with very dense phase coverage (over
2100 observations) of the low-mass eclipsing binary CU Cnc. In the light
curve analysis, carried out with the W-D program, we resolved an inherent
indeterminacy of the ratio of radii of the components (caused by the
partial eclipses) by fixing the luminosity ratio to the value obtained
from a high-resolution spectrum. Also, special care has been put in
modelling the conspicuous out-of-eclipse variations through surface
inhomogeities. The orbital and physical properties resulting from the
light curves were combined with the spectroscopic parameters from D99 to
achieve precisions better than 2\% in the absolute dimensions of CU Cnc
components.  As a result of this, CU Cnc is the third system (together
with YY Gem and CM Dra) to reach sufficient accuracy in its absolute
dimensions so that critical tests of stellar models can be performed.

To make the comparison with models more stringent, we made use of further
observational constraints. For example, the mean effective temperature of
CU Cnc was determined from multi-band photometry (optical through
infrared) and synthetic colours resulting from state-of-the-art atmosphere
models. Additionally, the space motions of CU Cnc strongly suggest
membership of the Castor moving group. This allowed us to infer rough
values for the age of the star ($\sim$320 Myr) and its chemical
composition ($Z\approx0.018$). With this host of observational information
-- quite uncommon for most eclipsing binaries --, we carried out a
detailed comparison with all of the recent low-mass stellar models. First,
the most reliable check is in the mass-radius diagram because these are
two quantities empirically determined from our analysis. The comparison
indicates that, whereas some of the models provide radius values that are
in reasonable agreement with the observations, the most sophisticated
models by Siess et al. (\cite{SFD97}) and Baraffe et al. (\cite{BCA98})
yield radii that are some 10\% smaller than observed. The study of three
additional observational planes indicates that CU Cnc's components are
also significantly cooler ($\sim$15\%) and fainter in the $V$ band
($\sim$1.5 mag) than predicted by most models.

A matter of some concern arises when comparing CU Cnc with other stars of
similar mass. The comparison indicates that CU Cnc is some 1.4 mag fainter
in the $V$ band and 0.35 mag fainter in the $K$ band. We have analysed a
number of possible scenarios to explain the discrepancy and ruled out most
of them on the basis of the available observational information. For
example, we have discarded the effect of stellar activity, an error in the
distance, a different age or an extreme chemical composition. Only two
scenarios appear to succeed in explaining the observed differences in the
mass-magnitude diagrams: {\em 1)} CU Cnc's components are, for a yet
unknown reason, about 10\% cooler than other stars of the same mass, and
{\em 2)} there is a certain amount of (circumstellar) dust absorption that
is responsible for the reddening of the spectral energy distribution. If
the latter turns out to be the correct scenario, this is a very exciting
result. Given the almost edge-on configuration of the eclipsing binary, we
could be observing CU Cnc through a coplanar dusty disk. Definite proof of
this should come from submillimeter observations where the thermal
emission of the dust could be detected. Note that CU Cnc belongs in the
same moving group and probably has similar age to \object{Vega} and
\object{Fomalhaut} (Barrado y Navascu\'es \cite{B98}), both stars with
bona-fide circumstellar dust disks (Holland et al. \cite{HGZ98}). With the
tentative detection of a Vega-like edge-on dust disk and its relatively
low mass and surface brightness, CU Cnc also becomes a very interesting
target for planet searches through radial velocity studies or the transit
method.

In our comparative study, we have concluded that the evolutionary models
of Swenson et al. (\cite{SFR94}) clearly stand out among all because they
achieve the best performance in reproducing all of the observed properties
of CU Cnc and even the relative position of the components (much better
constrained from the light curves). However, these models cannot obviously
fit the position of other 0.4-M$_{\sun}$ stars in the mass--absolute
magnitude diagrams, whereas models such as Siess et al. (\cite{SFD97}) and
especially Baraffe et al. (\cite{BCA98}) provide more reasonable fits.
Unfortunately, until the discrepancy between the radiative properties of
CU Cnc and stars of similar mass is resolved, no strong constraints on the
models can be set in the temperature and absolute magnitude diagrams. A
better determination of the parallax and an empirical estimation of the
metal abundance would surely help to clarify the situation.

Regardless of the issue with the temperature, the comparison of CU Cnc's
absolute dimensions (empirically-measured masses and radii) with stellar
model predictions clearly confirms the trend seen by TR02 for YY Gem and
V818 Tau. The systematic underestimation of the radii of the stars can
lead to severely underestimated stellar ages when using the H-R diagram
diagnostics. In spite of recent advances, it seems clear that further
adjustments are needed in evolution models for the lower main sequence to
achieve a good description of the observed stellar physical properties.

\begin{acknowledgements} 
Thanks are due to M. I. Andersen \& R. J. Irgens who kindly prepared and
performed CCD observations of CU Cnc with the Nordic Optical Telescope at
La Palma. I am most grateful to E. F. Guinan, G. Torres, A. Gim\'enez, C.
Jordi, J. V.  Clausen, D. Fern\'andez, F. Arenou, and R. Estalella for
their invaluable help and fruitful discussions during the course of this
work. The photoelectric observations in this publication have been
obtained with the Four College APT, which is supported by NSF grants
AST95-28506 and AST-0071260. The ING support astronomers are thanked for
carrying out the spectroscopic service observations used in this study.
The William Herschel Telescope is operated on the island of La Palma by
the Isaac Newton Group in the Spanish Observatorio del Roque de los
Muchachos of the Instituto de Astrofisica de Canarias. This publication
makes use of data products from the Two Micron All Sky Survey, which is a
joint project of the University of Massachusetts and the Infrared
Processing and Analysis Center/California Institute of Technology, funded
by the National Aeronautics and Space Administration and the National
Science Foundation. This research has made use of the SIMBAD database,
operated at CDS, Strasbourg, France. This research has made use of NASA's
Astrophysics Data System. 
\end{acknowledgements}

\end{document}